\documentclass{article}

\usepackage{times}
\usepackage{graphicx} 
\usepackage{subfigure} 
\usepackage{bm}
\usepackage{natbib}

\usepackage{algorithm}
\usepackage{algorithmic}

\usepackage{amsmath}

\usepackage{tabularx,booktabs}

\usepackage{hyperref}

\usepackage{iclr2017_conference} 

\usepackage{soul}

\renewcommand{\vec}{\mathbf}

\newcommand{\figref}[1]{Fig.~\ref{#1}}
\newcommand{\tblref}[1]{Table~\ref{#1}}

\newcommand{\eqnref}[1]{Eq.~\eqref{#1}}

\usepackage{xspace}

\newcommand*{\eg}{e.g.\@\xspace}

\bmdefine{\bO}{O}
\bmdefine{\bC}{C}
\bmdefine{\bc}{c}
\def\bo{\vec{o}}
\bmdefine{\bW}{W}
\bmdefine{\bmu}{\mu}
\bmdefine{\bQ}{Q}
\bmdefine{\bq}{q}
\bmdefine{\bw}{w}
\bmdefine{\bU}{U}
\bmdefine{\bL}{L}
\bmdefine{\bu}{u}
\bmdefine{\bZero}{0}
\bmdefine{\bI}{I}
\bmdefine{\bR}{R}
\bmdefine{\bP}{P}
\bmdefine{\br}{r}
\bmdefine{\bmm}{m}
\bmdefine{\bsigma}{\sigma}
\bmdefine{\bSigma}{\Sigma}
\bmdefine{\bOmega}{\Omega}
\bmdefine{\bomega}{\omega}
\bmdefine{\bS}{S}
\bmdefine{\bA}{A}
\bmdefine{\bC}{C}
\bmdefine{\bM}{M}
\bmdefine{\bg}{g}
\bmdefine{\bs}{s}
\bmdefine{\bpsi}{\psi}
\bmdefine{\bPsi}{\Psi}
\bmdefine{\bphi}{\phi}
\bmdefine{\bPhi}{\Phi}
\bmdefine{\bPi}{\Pi}
\bmdefine{\bpi}{\pi}
\bmdefine{\bLambda}{\Lambda}
\bmdefine{\blambda}{\lambda}
\bmdefine{\bB}{B}
\bmdefine{\bb}{b}
\def\bl{\vec{l}}
\bmdefine{\bd}{d}
\bmdefine{\bD}{D}
\bmdefine{\bY}{Y}
\bmdefine{\bG}{G}
\bmdefine{\bp}{p}
\bmdefine{\bxi}{\xi}
\bmdefine{\bmeta}{\eta}
\bmdefine{\bzeta}{\zeta}
\bmdefine{\bk}{k}
\bmdefine{\bK}{K}
\bmdefine{\bF}{F}
\bmdefine{\bv}{v}
\bmdefine{\bX}{X}
\def\bx{\vec{x}}
\bmdefine{\by}{y}
\bmdefine{\bz}{z}
\bmdefine{\bZ}{Z}
\bmdefine{\bcalX}{\mathcal{X}}
\bmdefine{\bH}{H}
\bmdefine{\bh}{h}
\bmdefine{\bcalH}{\mathcal{H}}
\bmdefine{\bV}{V}

\def\Gauss{\mathcal{N}}

\newcommand{\argmax}{\operatornamewithlimits{\mathrm{arg\,max}}}

\definecolor {GoogleRed}   {rgb}{0.97265625, 0.00390625, 0.00390625}
\definecolor {GoogleBlue}  {rgb}{0.0078125,  0.3984375,  0.78125}
\definecolor {GoogleYellow}{rgb}{0.9453125,  0.70703125, 0.05859375}
\definecolor {GoogleGreen} {rgb}{0.0,        0.57421875, 0.23046875}

\newcolumntype{L}{>{\raggedright\arraybackslash}X}
\newcolumntype{C}{>{\centering\arraybackslash}X}
\newcolumntype{R}{>{\raggedleft\arraybackslash}X}

\title{
WaveNet: A Generative Model for Raw Audio
}

\author{A\"{a}ron~van~den~Oord \And
\hspace{-0.5cm}Sander~Dieleman \And
Heiga~Zen$^\dagger$ \AND
Karen~Simonyan \And
Oriol~Vinyals \And
Alex~Graves \AND
Nal~Kalchbrenner \And
\hspace{1cm}Andrew~Senior \And
Koray~Kavukcuoglu \AND \\
\{avdnoord, sedielem, heigazen, simonyan, vinyals, gravesa, nalk, andrewsenior, korayk\}@google.com\\
Google DeepMind, London, UK \\ 
$^\dagger$\,Google, London, UK
}

\begin{document}

\maketitle

\vskip 0.3in

\begin{abstract}
This paper introduces WaveNet, a deep neural network for generating raw audio waveforms. The model is fully probabilistic and autoregressive, with the predictive distribution for each audio sample conditioned on all previous ones; nonetheless we show that it can be efficiently trained on data with tens of thousands of samples per second of audio. 
When applied to text-to-speech, it yields state-of-the-art performance, with human listeners rating it as significantly more natural sounding than the best parametric and concatenative systems for both English and Mandarin. A single WaveNet can capture the characteristics of many different speakers with equal fidelity, and can switch between them by conditioning on the speaker identity. When trained to model music, we find that it generates novel and often highly realistic musical fragments. We also show that it can be employed as a discriminative model, returning promising results for phoneme recognition.
\end{abstract}

\section{Introduction}

This work explores raw audio generation techniques, inspired by recent advances in neural autoregressive generative models that model complex distributions such as images \citep{van2016pixel, ConditionalPixelCNN} and text \citep{RafalLanguage}. Modeling joint probabilities over pixels or words using neural architectures as products of conditional distributions yields state-of-the-art generation.

\begin{figure}[htb]
\includegraphics[height=36pt,width=\linewidth]{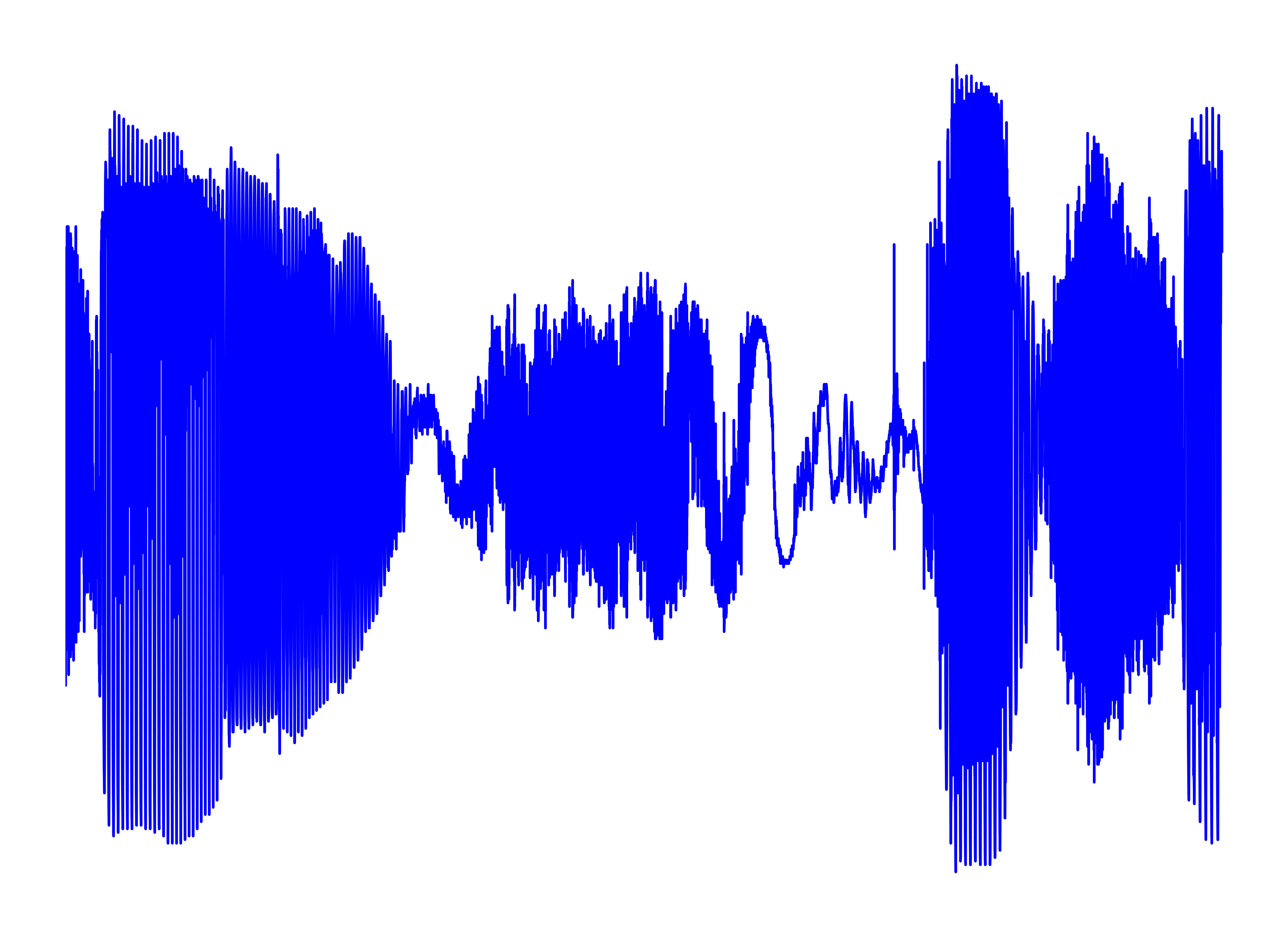}
\caption{A second of generated speech.}
\label{fig:audio}
\end{figure}

Remarkably, these architectures are able to model distributions over thousands of random variables (\eg 64$\times$64 pixels as in PixelRNN \citep{van2016pixel}). The question this paper addresses is whether similar approaches can succeed in generating wideband raw audio waveforms, which are signals with very high temporal resolution, at least 16,000 samples per second (see \figref{fig:audio}).

This paper introduces \emph{WaveNet}, an audio generative model based on the PixelCNN \citep{van2016pixel, ConditionalPixelCNN} architecture. The main contributions of this work are as follows:
\begin{itemize}
    \item We show that WaveNets can generate raw speech signals with subjective naturalness never before reported in the field of text-to-speech (TTS), as assessed by human raters.
    \item In order to deal with long-range temporal dependencies needed for raw audio generation, we develop new architectures based on dilated causal convolutions, which exhibit very large receptive fields.
    \item We show that when conditioned on a speaker identity, a single model can be used to generate different voices.
    \item The same architecture shows strong results when tested on a small speech recognition dataset, and is promising when used to generate other audio modalities such as music.
\end{itemize}

We believe that WaveNets provide a generic and flexible framework for tackling many applications that rely on audio generation (\eg TTS, music, speech enhancement, voice conversion, source separation).

\section{WaveNet}
\label{sec:wavenet}
In this paper we introduce a new generative model operating directly on the raw audio waveform.  
The joint probability of a waveform $\vec{x} = \{ x_1, \dots, x_T \}$ is factorised as a product of conditional probabilities as follows:
\begin{equation}
p\left(\vec{x}\right) = \prod_{t=1}^{T} p\left(x_t \mid x_1, \dots ,x_{t-1}\right)
\label{eq:px}
\end{equation}
Each audio sample $x_t$ is therefore conditioned on the samples at all previous timesteps.

Similarly to PixelCNNs \citep{van2016pixel, ConditionalPixelCNN}, the conditional probability distribution is modelled by a stack of convolutional layers. There are no pooling layers in the network, and the output of the model has the same time dimensionality as the input. The model outputs a categorical distribution over the next value $x_{t}$ with a softmax layer and it is optimized to maximize the log-likelihood of the data w.r.t. the parameters. Because log-likelihoods are tractable, we tune hyper-parameters on a validation set and can easily measure if the model is overfitting or underfitting.

\subsection{Dilated Causal Convolutions}

\begin{figure}[h]
\centering
\includegraphics[width=0.85\linewidth]{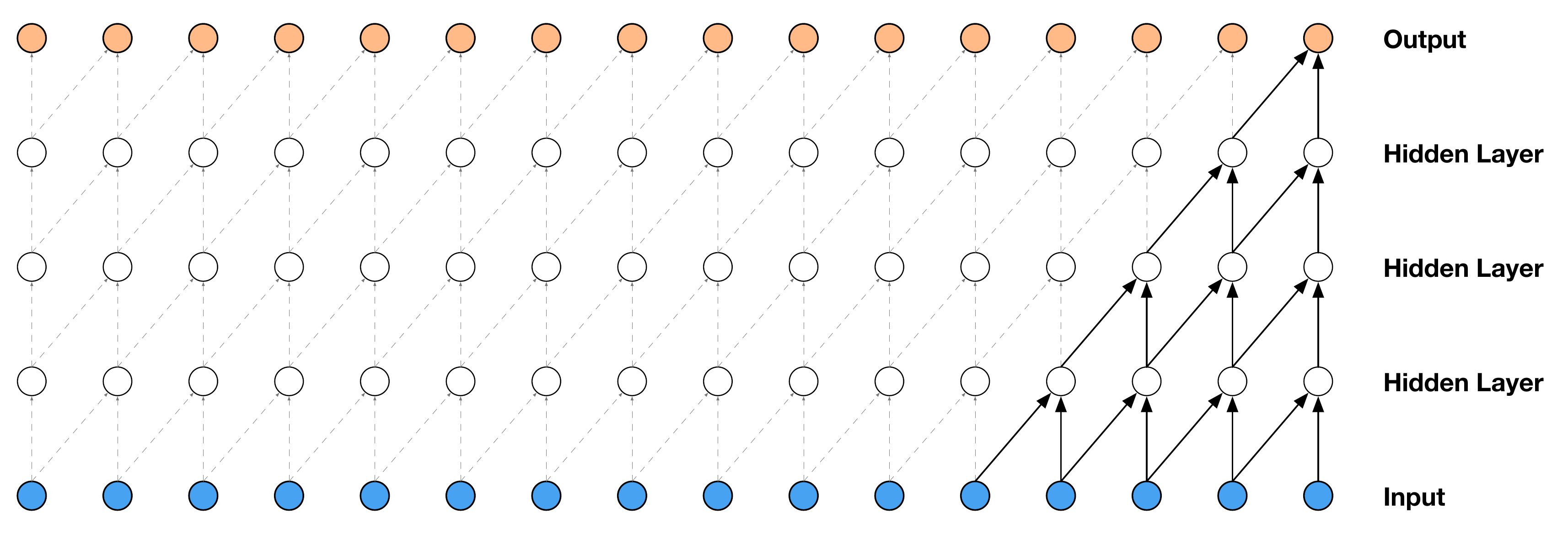}
\caption{Visualization of a stack of causal convolutional layers.}
\label{fig:masked_convolution}
\end{figure}

The main ingredient of WaveNet are causal convolutions. By using causal convolutions, we make sure the model cannot violate the ordering in which we model the data: the prediction $p\left(x_{t+1} \mid x_1,...,x_{t}\right)$ emitted by the model at timestep $t$ cannot depend on any of the future timesteps $x_{t+1}, x_{t+2},\dots,x_T$ as shown in \figref{fig:masked_convolution}. For images, the equivalent of a causal convolution is a masked convolution \citep{van2016pixel} which can be implemented by constructing a mask tensor and doing an elementwise multiplication of this mask with the convolution kernel before applying it. For 1-D data such as audio one can more easily implement this by shifting the output of a normal convolution by a few timesteps. 

At training time, the conditional predictions for all timesteps can be made in parallel because all timesteps of ground truth $\vec{x}$ are known. When generating with the model, the predictions are sequential: after each sample is predicted, it is fed back into the network to predict the next sample.

Because models with causal convolutions do not have recurrent connections, they are typically faster to train than RNNs, especially when applied to very long sequences. One of the problems of causal convolutions is that they require many layers, or large filters to increase the receptive field. For example, in \figref{fig:masked_convolution} the receptive field is only 5 (= \#layers + filter length - 1). In this paper we use dilated convolutions to increase the receptive field by orders of magnitude, without greatly increasing computational cost. 

\begin{figure}[ht]
\centering
\includegraphics[trim={6.625in 0 4.95in 0},clip,width=0.85\linewidth]{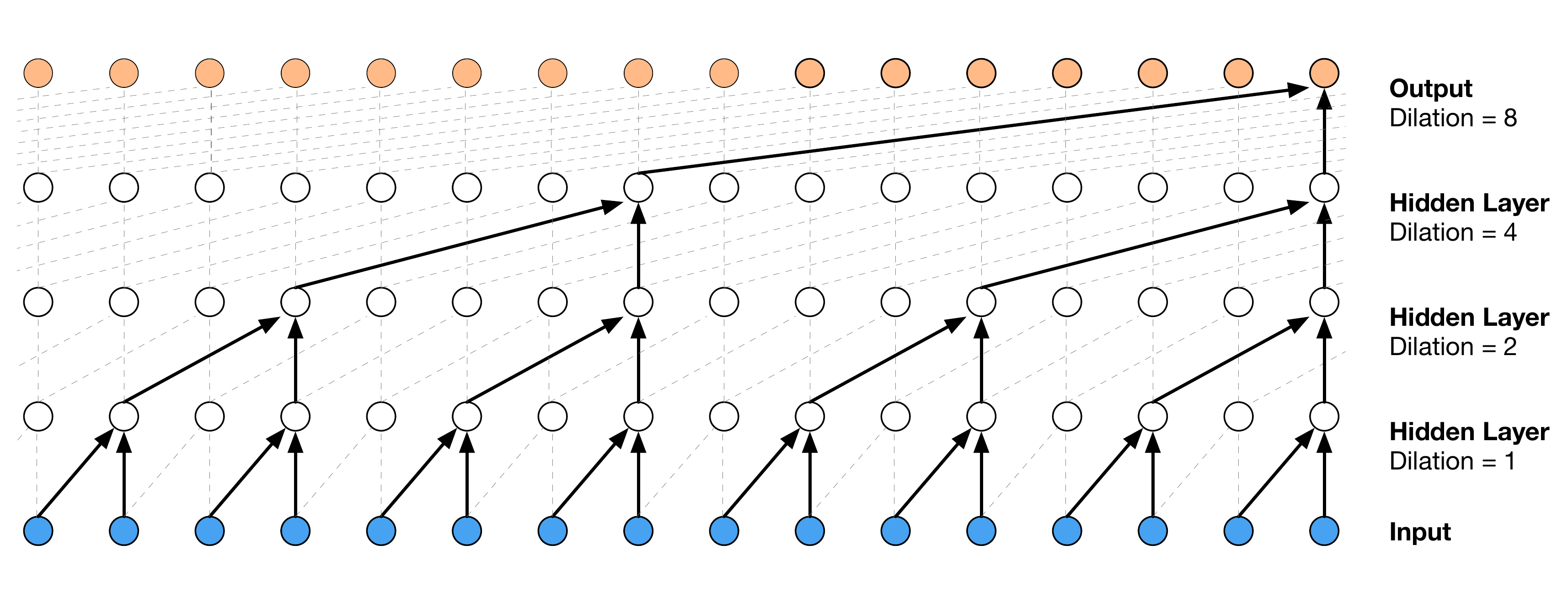}
\caption{Visualization of a stack of \emph{dilated} causal convolutional layers.}
\label{fig:masked_dilated_convolution}
\end{figure}

A dilated convolution (also called \emph{\`a trous}, or convolution with holes) is a convolution where the filter is applied over an area larger than its length by skipping input values with a certain step. It is equivalent to a convolution with a larger filter derived from the original filter by dilating it with zeros, but is significantly more efficient. A dilated convolution effectively allows the network to operate on a coarser scale than with a normal convolution. This is similar to pooling or strided convolutions, but here the output has the same size as the input. As a special case, dilated convolution with dilation $1$ yields the standard convolution.  \figref{fig:masked_dilated_convolution} depicts dilated causal convolutions for dilations $1$, $2$, $4$, and $8$. Dilated convolutions have previously been used in various contexts, \eg signal processing \citep{Holschneider1989,Dutilleux1989}, and image segmentation \citep{chen14semantic,YuKoltun2016}.

Stacked dilated convolutions enable networks to have very large receptive fields with just a few layers, while preserving the input resolution throughout the network as well as computational efficiency. In this paper, the dilation is doubled for every layer up to a limit and then repeated: \eg $$1,2,4,\dots,512,1,2,4,\dots,512,1,2,4,\dots,512.$$
The intuition behind this configuration is two-fold. First, exponentially increasing the dilation factor results in exponential receptive field growth with depth~\citep{YuKoltun2016}.  For example each $1,2,4,\dots,512$ block has receptive field of size $1024$, and can be seen as a more efficient and discriminative (non-linear) counterpart of a $1\times1024$ convolution. Second, stacking these blocks further increases the model capacity and the receptive field size.

\subsection{Softmax distributions}

One approach to modeling the conditional distributions $p\left(x_t \mid x_1, \dots ,x_{t-1}\right)$ over the individual audio samples would be to use a mixture model such as a mixture density network \citep{MDN} or mixture of conditional Gaussian scale mixtures (MCGSM) \citep{theis2015generative}.
However, \cite{van2016pixel} showed that a softmax distribution tends to work better, even when the data is implicitly continuous (as is the case for image pixel intensities or audio sample values). One of the reasons is that a categorical distribution is more flexible and can more easily model arbitrary distributions because it makes no assumptions about their shape.

Because raw audio is typically stored as a sequence of 16-bit integer values (one per timestep), a softmax layer would need to output 65,536 probabilities per timestep to model all possible values. To make this more tractable, we first apply a $\mu$-law companding transformation \citep{G711} to the data, and then quantize it to 256 possible values:
$$
f\left(x_t\right) = \operatorname{sign}(x_t) \frac{\ln \left(1+\mu \left| x_t \right|\right)}{\ln \left(1+\mu\right)},
$$
where $-1 < x_t < 1$ and $\mu = 255$. 
This non-linear quantization produces a significantly better reconstruction than a simple linear quantization scheme. Especially for speech, we found that the reconstructed signal after quantization sounded very similar to the original.

\subsection{Gated activation units}

We use the same gated activation unit as used in the gated PixelCNN \citep{ConditionalPixelCNN}:
\begin{equation}
\vec{z} = \tanh \left(W_{f, k} \ast \vec{x}\right) \odot \sigma \left(W_{g, k} \ast \vec{x} \right), \label{eq:gated_activation}
\end{equation}
where $\ast$ denotes a convolution operator, $\odot$ denotes an element-wise multiplication operator, $\sigma(\cdot)$ is a sigmoid function, $k$ is the layer index, $f$ and $g$ denote filter and gate, respectively, and $W$ is a learnable convolution filter.
In our initial experiments, we observed that this non-linearity worked significantly better than the rectified linear activation function \citep{nair2010rectified} for modeling audio signals.

\subsection{Residual and skip connections}

\begin{figure}[h]
\centering
\includegraphics[width=0.75\linewidth]{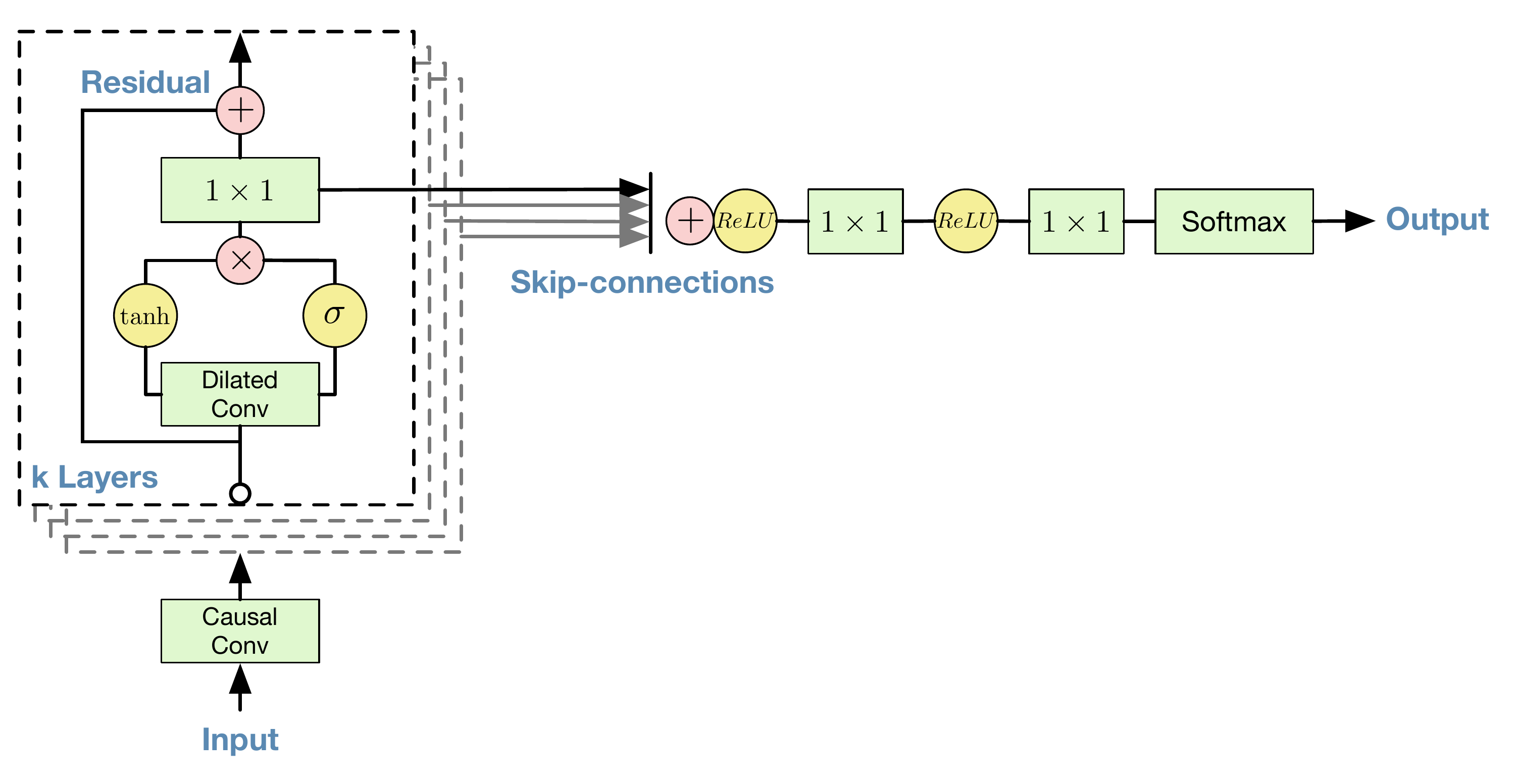}
\caption{Overview of the residual block and the entire architecture.}
\label{fig:architecture}
\end{figure}

Both residual \citep{he15deep} and parameterised skip connections are used throughout the network, to speed up convergence and enable training of much deeper models. In \figref{fig:architecture} we show a residual block of our model, which is stacked many times in the network.

\subsection{Conditional WaveNets}

Given an additional input $\vec{h}$, WaveNets can model the conditional distribution $p\left(\vec{x} \mid \vec{h}\right)$ of the audio given this input. \eqnref{eq:px} now becomes
\begin{equation}
p\left( \vec{x} \mid \vec{h} \right) = \prod_{t=1}^{T} p\left(x_t \mid x_1, \dots ,x_{t-1}, \vec{h}\right).
\label{eq:pxh}
\end{equation}

By conditioning the model on other input variables, we can guide WaveNet's generation to produce audio with the required characteristics. For example, in a multi-speaker setting we can choose the speaker by feeding the speaker identity to the model as an extra input. Similarly, for TTS we need to feed information about the text as an extra input.

We condition the model on other inputs in two different ways: global conditioning and local conditioning. Global conditioning is characterised by a single latent representation $\vec{h}$ that influences the output distribution across all timesteps, \eg a speaker embedding in a TTS model. The activation function from \eqnref{eq:gated_activation} now becomes:
$$
\vec{z} = \tanh \left(W_{f, k} \ast \vec{x} + V_{f, k}^T \vec{h}\right) \odot \sigma \left(W_{g, k} \ast \vec{x} + V_{g, k}^T \vec{h} \right). \label{eq:global_conditioning}
$$
where $V_{*, k}$ is a learnable linear projection, and the vector $V_{*, k}^T\vec{h}$ is broadcast over the time dimension.

For local conditioning we have a second timeseries $h_t$, possibly with a lower sampling frequency than the audio signal, \eg linguistic features in a TTS model. We first transform this time series using a transposed convolutional network (learned upsampling) that maps it to a new time series $\vec{y} = f(\vec{h})$ with the same resolution as the audio signal, which is then used in the activation unit as follows:
$$
\vec{z} = \tanh \left(W_{f, k} \ast \vec{x} + V_{f, k} \ast \vec{y}\right) \odot \sigma \left(W_{g, k} \ast \vec{x} + V_{g, k} \ast \vec{y}\right), \label{eq:local_conditioning}
$$
where $V_{f, k} \ast \vec{y}$ is now a $1\times 1$ convolution. As an alternative to the transposed convolutional network, it is also possible to use $V_{f, k} \ast \vec{h}$ and repeat these values across time. We saw that this worked slightly worse in our experiments.

\subsection{Context stacks}

We have already mentioned several different ways to increase the receptive field size of a WaveNet: increasing the number of dilation stages, using more layers, larger filters, greater dilation factors, or a combination thereof. A complementary approach is to use a separate, smaller \emph{context} stack that processes a long part of the audio signal and locally conditions a larger WaveNet that processes only a smaller part of the audio signal (cropped at the end). One can use multiple context stacks with varying lengths and numbers of hidden units. Stacks with larger receptive fields have fewer units per layer. Context stacks can also have pooling layers to run at a lower frequency. This keeps the computational requirements at a reasonable level and is consistent with the intuition that less capacity is required to model temporal correlations at longer timescales.

\section{Experiments}

To measure WaveNet's audio modelling performance, we evaluate it on three different tasks: multi-speaker speech generation (not conditioned on text), TTS, and music audio modelling. We provide samples drawn from WaveNet for these experiments on the accompanying webpage:\\ \url{https://www.deepmind.com/blog/wavenet-generative-model-raw-audio/}.

\subsection{Multi-Speaker Speech Generation}

For the first experiment we looked at free-form speech generation (not conditioned on text). We used the English multi-speaker corpus from CSTR voice cloning toolkit (VCTK) \citep{VCTK} and conditioned WaveNet only on the speaker. The conditioning was applied by feeding the speaker ID to the model in the form of a one-hot vector. The dataset consisted of 44 hours of data from 109 different speakers.

Because the model is not conditioned on text, it generates non-existent but human language-like words in a smooth way with realistic sounding intonations. This is similar to generative models of language or images, where samples look realistic at first glance, but are clearly unnatural upon closer inspection. The lack of long range coherence is partly due to the limited size of the model's receptive field (about 300 milliseconds), which means it can only remember the last 2--3 phonemes it produced.

A single WaveNet was able to model speech from any of the speakers by conditioning it on a one-hot encoding of a speaker. This confirms that it is powerful enough to capture the characteristics of all 109 speakers from the dataset in a single model. We observed that adding speakers resulted in better validation set performance compared to training solely on a single speaker. This suggests that WaveNet's internal representation was shared among multiple speakers.

Finally, we observed that the model also picked up on other characteristics in the audio apart from the voice itself. For instance, it also mimicked the acoustics and recording quality, as well as the breathing and mouth movements of the speakers.

\subsection{Text-To-Speech}
For the second experiment we looked at TTS.  
We used the same single-speaker speech databases from which Google's North American English and Mandarin Chinese TTS systems are built.
The North American English dataset contains 24.6 hours of speech data, and the Mandarin Chinese dataset contains 34.8 hours; both were spoken by professional female speakers.

WaveNets for the TTS task were locally conditioned on \emph{linguistic features} which were derived from input texts.
We also trained WaveNets conditioned on the logarithmic fundamental frequency ($\log F_0$) values in addition to the linguistic features.
External models predicting $\log F_0$ values and phone durations from linguistic features were also trained for each language.
The receptive field size of the WaveNets was 240 milliseconds.
As example-based and model-based speech synthesis baselines, hidden Markov model (HMM)-driven unit selection concatenative \citep{Xavi_Barracuda_interspeech}
and long short-term memory recurrent neural network (LSTM-RNN)-based statistical parametric \citep{Zen_LSTMprod_Interspeech} speech synthesizers were built.
Since the same datasets and linguistic features were used to train both the baselines and WaveNets, these speech synthesizers could be fairly compared.

To evaluate the performance of WaveNets for the TTS task, subjective paired comparison tests and mean opinion score (MOS) tests were conducted.
In the paired comparison tests, after listening to each pair of samples, the subjects were asked to choose which they preferred, though they could choose ``neutral'' if they did not have any preference.
In the MOS tests, after listening to each stimulus, the subjects were asked to rate the naturalness of the stimulus in a five-point Likert scale score (1: Bad, 2: Poor, 3: Fair, 4: Good, 5: Excellent).
Please refer to Appendix~\ref{appendix:tts_experiment} for details.

\figref{fig:sxs2} shows a selection of the subjective paired comparison test results (see  Appendix~\ref{appendix:tts_experiment} for the complete table).
It can be seen from the results that WaveNet outperformed the baseline statistical parametric and concatenative speech synthesizers in both languages.
We found that WaveNet conditioned on linguistic features could synthesize speech samples with natural segmental quality but sometimes it had unnatural prosody by stressing wrong words in a sentence.
This could be due to the long-term dependency of $F_0$ contours: the size of the receptive field of the WaveNet, 240 milliseconds, was not long enough to capture such long-term dependency.  
WaveNet conditioned on both linguistic features and $F_0$ values did not have this problem: the external $F_0$ prediction model runs at a lower frequency (200 Hz) so it can learn long-range dependencies that exist in $F_0$ contours.
 
\begin{figure}[htbp]
\centering
\includegraphics[width=0.65\linewidth]{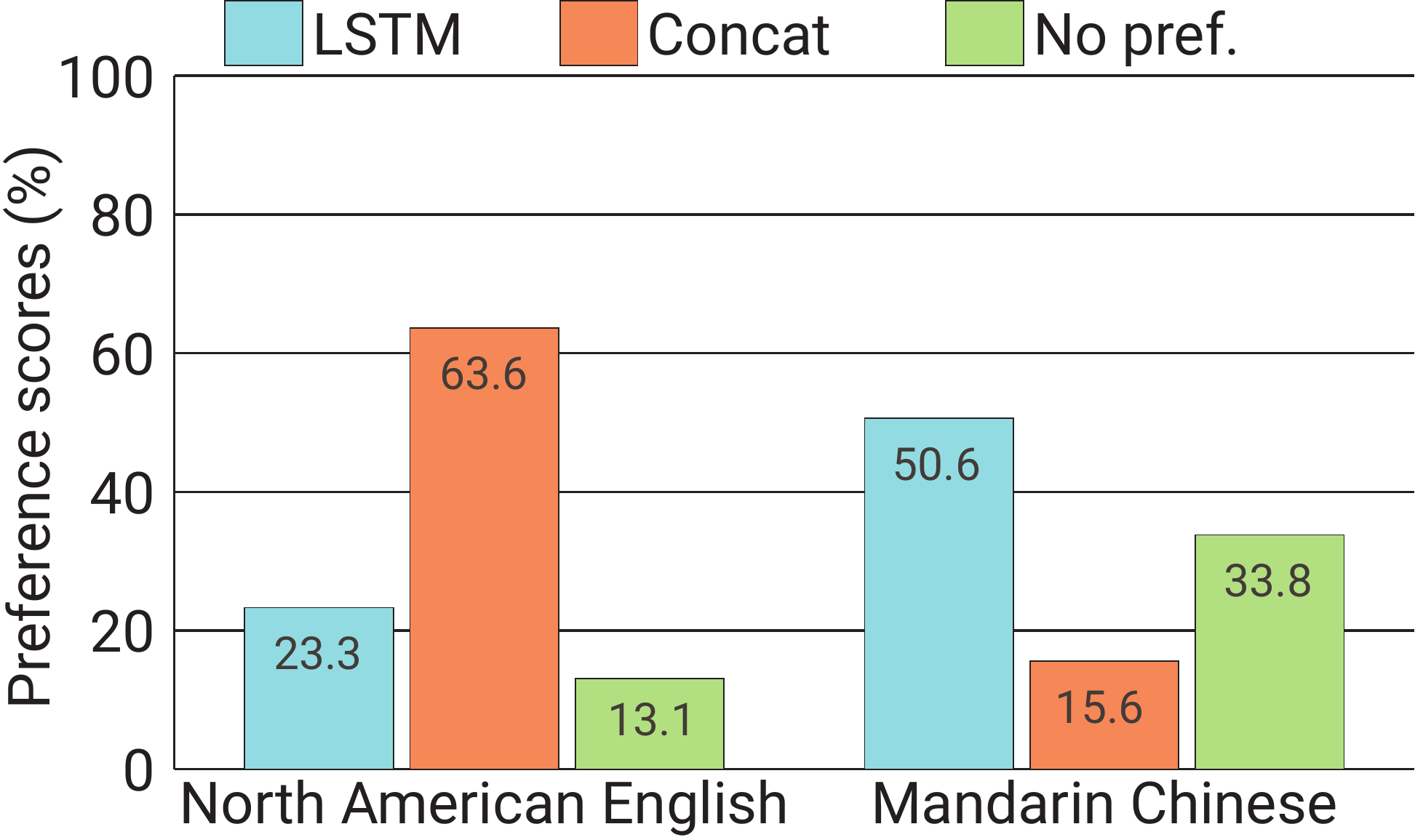}%
\vspace{8mm}\par
\includegraphics[width=0.65\linewidth]{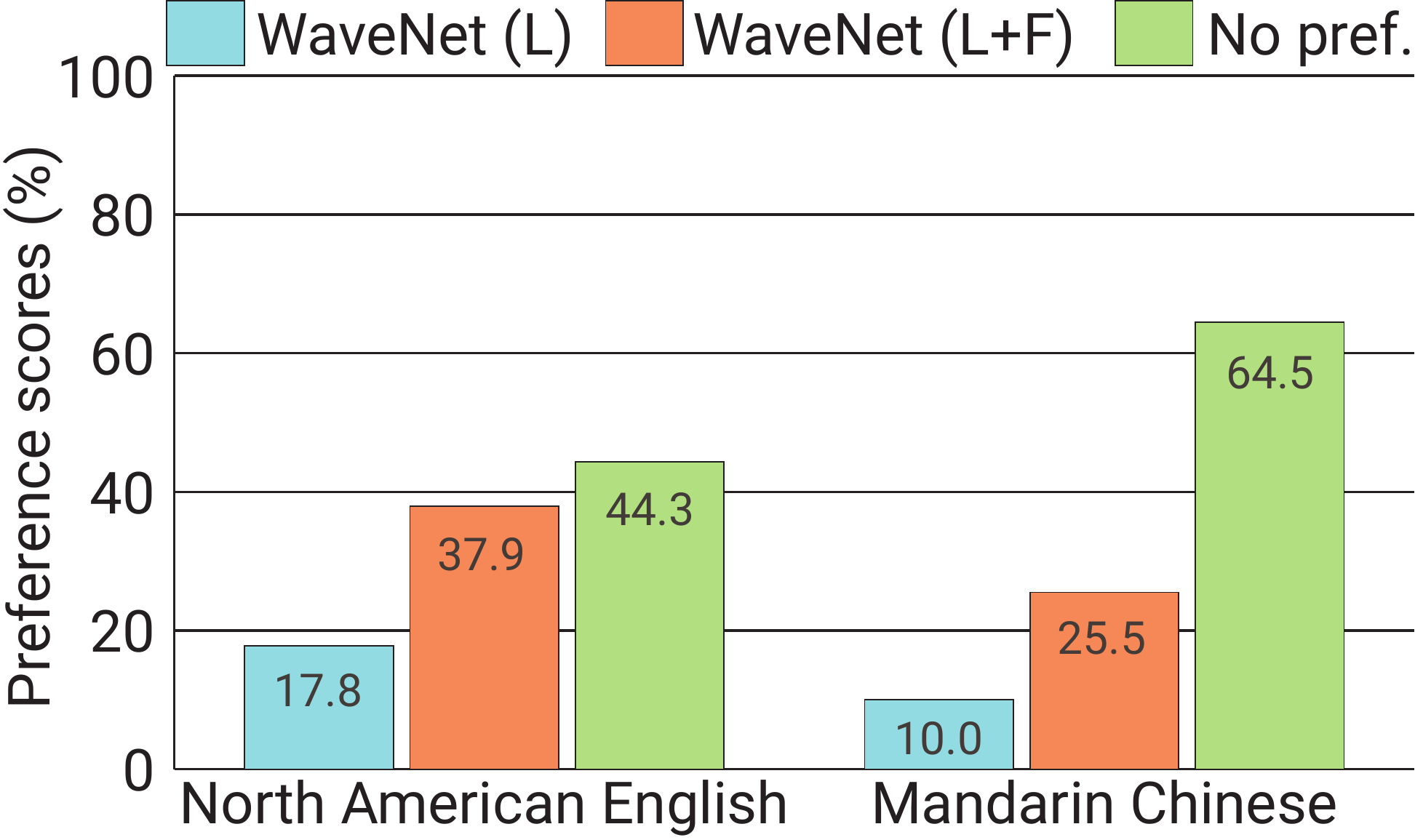}%
\vspace{8mm}\par
\includegraphics[width=0.65\linewidth]{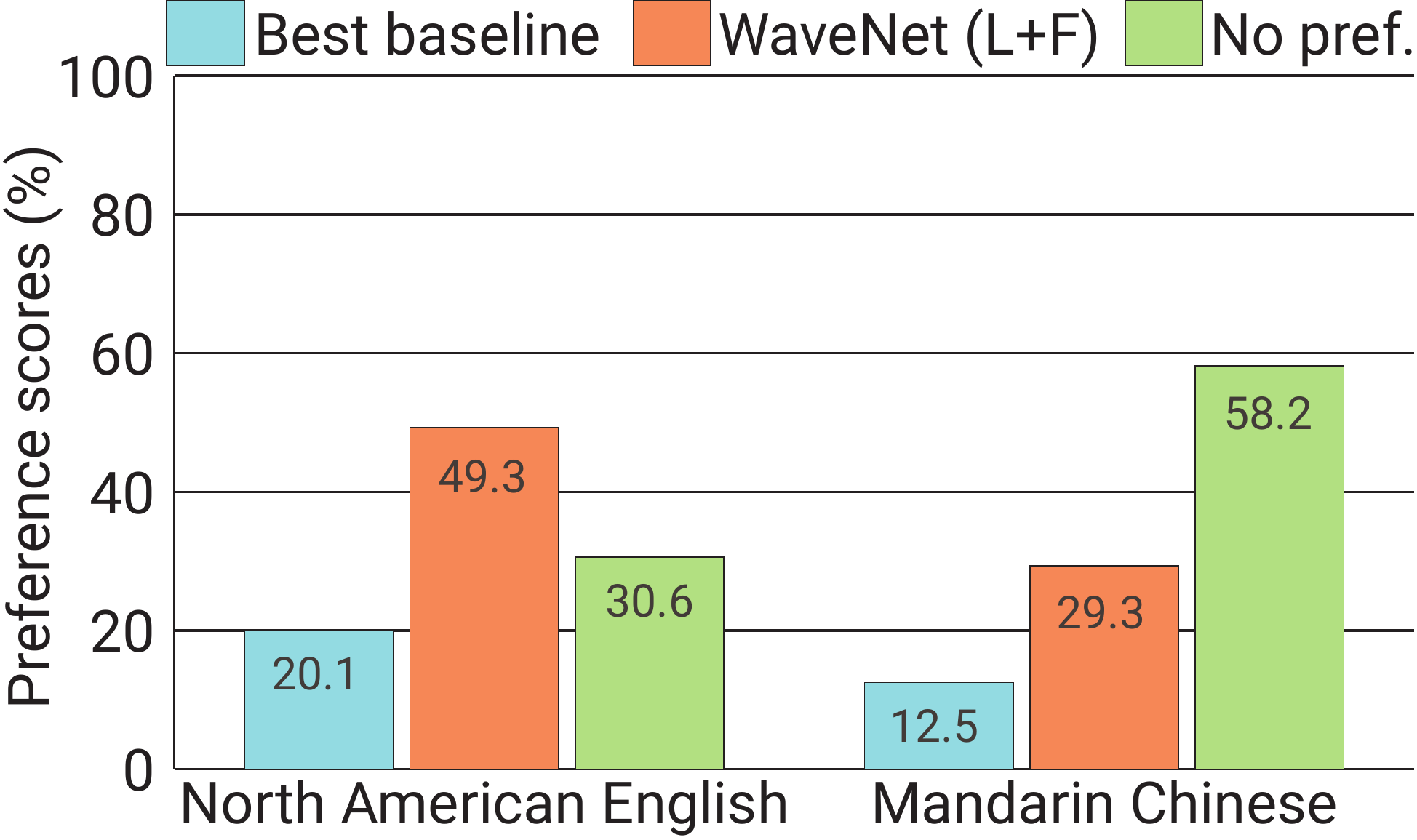}%
\caption{Subjective preference scores (\%) of speech samples between (top) two baselines, (middle) two WaveNets, and (bottom) the best baseline and WaveNet.  Note that \textsf{LSTM} and 
\textsf{Concat} correspond to LSTM-RNN-based statistical parametric and HMM-driven unit selection concatenative baseline synthesizers, and \textsf{WaveNet (L)} and \textsf{WaveNet (L+F)} correspond to the WaveNet conditioned on linguistic features only and that conditioned on both linguistic features and $\log F_0$ values.}
\label{fig:sxs2}
\end{figure}

\tblref{tab:mos} show the MOS test results.
It can be seen from the table that WaveNets achieved 5-scale MOSs in naturalness above 4.0, which were significantly better than those from the baseline systems.
They were the highest ever reported MOS values with these training datasets and test sentences.
The gap in the MOSs from the best synthetic speech to the natural ones decreased from 0.69 to 0.34 (51\%) in US English and 0.42 to 0.13 (69\%) in Mandarin Chinese.

\begin{table}[htbp]
  \centering
  \begin{tabularx}{0.94\textwidth}{L|CC}
    \toprule
    & \multicolumn{2}{|c}{\textbf{Subjective 5-scale MOS in naturalness}} \\ \cmidrule{2-3}
    \textbf{Speech samples} & North American English & Mandarin Chinese \\ \midrule\midrule
    LSTM-RNN parametric  & 3.67 $\pm$ 0.098 & 3.79 $\pm$ 0.084 \\ 
    HMM-driven concatenative & 3.86 $\pm$ 0.137 & 3.47 $\pm$ 0.108  \\
    \textbf{WaveNet} (L+F) & \textbf{4.21} $\pm$ 0.081 & \textbf{4.08} $\pm$ 0.085 \\
    \midrule
    Natural (8-bit $\mu$-law) & 4.46 $\pm$ 0.067 & 4.25 $\pm$ 0.082 \\
    Natural (16-bit linear PCM) & 4.55 $\pm$ 0.075 & 4.21 $\pm$ 0.071 \\
    \bottomrule
  \end{tabularx}%
  \caption{Subjective 5-scale mean opinion scores of speech samples from  LSTM-RNN-based statistical parametric, HMM-driven unit selection concatenative, and proposed WaveNet-based speech synthesizers, 8-bit $\mu$-law encoded natural speech, and 16-bit linear pulse-code modulation (PCM) natural speech. WaveNet improved the previous state of the art significantly, reducing the gap between natural speech and best previous model by more than 50\%.}
  \label{tab:mos}
\end{table}%

\subsection{Music}

For out third set of experiments we trained WaveNets to model two music datasets:
\begin{itemize}
    \item the MagnaTagATune dataset \citep{law2009input}, which consists of about 200 hours of music audio. Each 29-second clip is annotated with tags from a set of 188, which describe the genre, instrumentation, tempo, volume and mood of the music.
    \item the YouTube piano dataset, which consists of about 60 hours of solo piano music obtained from YouTube videos. Because it is constrained to a single instrument, it is considerably easier to model.
\end{itemize}
Although it is difficult to quantitatively evaluate these models, a subjective evaluation is possible by listening to the samples they produce.
We found that enlarging the receptive field was crucial to obtain samples that sounded musical. Even with a receptive field of several seconds, the models did not enforce long-range consistency which resulted in second-to-second variations in genre, instrumentation, volume and sound quality. Nevertheless, the samples were often harmonic and aesthetically pleasing, even when produced by unconditional models.

Of particular interest are conditional music models, which can generate music given a set of tags specifying e.g. genre or instruments. Similarly to conditional speech models, we insert biases that depend on a binary vector representation of the tags associated with each training clip. This makes it possible to control various aspects of the output of the model when sampling, by feeding in a binary vector that encodes the desired properties of the samples. We have trained such models on the MagnaTagATune dataset; although the tag data bundled with the dataset was relatively noisy and had many omissions, after cleaning it up by merging similar tags and removing those with too few associated clips, we found this approach to work reasonably well.

\subsection{Speech Recognition}

Although WaveNet was designed as a generative model, it can straightforwardly be adapted to discriminative audio tasks such as speech recognition. 

Traditionally, speech recognition research has largely focused on using log mel-filterbank energies or mel-frequency cepstral coefficients (MFCCs), but has been moving to raw audio recently \citep{palaz2013estimating,tuske2014acoustic,hoshen2015speech,sainath2015learning}. Recurrent neural networks such as LSTM-RNNs \citep{LSTM} have been a key component in these new speech classification pipelines, because they allow for building models with long range contexts. With WaveNets we have shown that layers of dilated convolutions allow the receptive field to grow longer in a much cheaper way than using LSTM units.

As a last experiment we looked at speech recognition with WaveNets on the TIMIT \citep{TIMIT} dataset. For this task we added a mean-pooling layer after the dilated convolutions that aggregated the activations to coarser frames spanning 10 milliseconds (160$\times$ downsampling). The pooling layer was followed by a few non-causal convolutions. We trained WaveNet with two loss terms, one to predict the next sample and one to classify the frame, the model generalized better than with a single loss and achieved $18.8$ PER on the test set, which is to our knowledge the best score obtained from a model trained directly on raw audio on TIMIT.

\section{Conclusion}
This paper has presented WaveNet, a deep generative model of audio data that operates directly at the waveform level. WaveNets are autoregressive and combine causal filters with dilated convolutions to allow their receptive fields to grow exponentially with depth, which is important to model the long-range temporal dependencies in audio signals. We have shown how WaveNets can be conditioned on other inputs in a global (\eg speaker identity) or local way (\eg linguistic features). When applied to TTS, WaveNets produced samples that outperform the current best TTS systems in subjective naturalness. Finally, WaveNets showed very promising results when applied to music audio modeling and speech recognition.

\section*{Acknowledgements} 
The authors would like to thank Lasse Espeholt, Jeffrey De Fauw and Grzegorz Swirszcz for their inputs, Adam Cain, Max Cant and Adrian Bolton for their help with artwork, Helen King, Steven Gaffney and Steve Crossan for helping to manage the project, Faith Mackinder for help  with preparing the blogpost, James Besley for legal support and Demis Hassabis for managing the project and his inputs.
\bibliography{main,references-short}
\bibliographystyle{icml2016}

\appendix 

\section{Text-to-Speech Background}
The goal of TTS synthesis is to render  naturally sounding speech signals given a text to be synthesized.
Human speech production process first translates a text (or concept) into movements of muscles associated with
articulators and speech production-related organs.
Then using air-flow from lung, vocal source excitation signals, which contain both periodic
(by vocal cord vibration) and aperiodic (by turbulent noise) components, are generated.
By filtering the vocal source excitation signals by time-varying vocal tract transfer functions
controlled by the articulators, their frequency characteristics are modulated.
Finally, the generated speech signals are emitted.
The aim of TTS is to mimic this process by computers in some way.

TTS can be viewed as a sequence-to-sequence mapping problem;
from a sequence of 
discrete symbols (text) to a real-valued time series (speech signals).
A typical TTS pipeline has two parts; 1) text analysis and 2) speech synthesis.
The text analysis part typically includes a number of natural language processing (NLP) steps,
such as sentence segmentation, word segmentation, text normalization,
part-of-speech (POS) tagging,
and grapheme-to-phoneme (G2P) conversion.
It takes a word sequence as input and outputs
a phoneme sequence with a variety of linguistic contexts.
The speech synthesis part takes the context-dependent phoneme sequence as its input and outputs a synthesized speech waveform.
This part typically includes prosody prediction and speech waveform generation.

\begin{figure}[!t]
  \centering
  \includegraphics[width=0.6\textwidth]{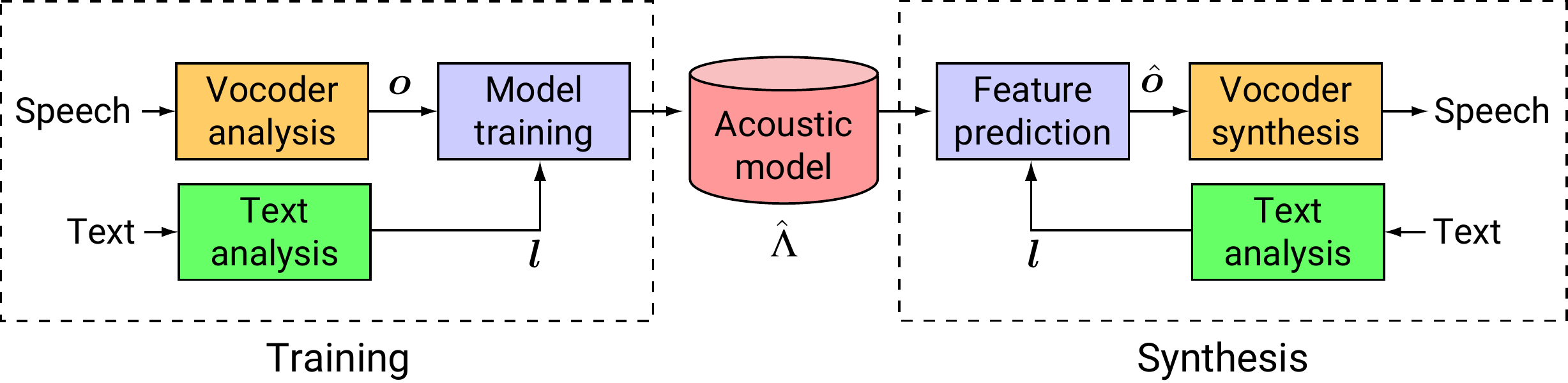}
  \caption{Outline of statistical parametric speech synthesis.}
  \label{fig:outline_spss}
\end{figure}

There are two main approaches to realize the speech synthesis part; non-parametric, example-based approach known as concatenative speech synthesis \citep{PSOLA,ATR_nutalk,Hunt_UnitSelection_ICASSP}, and parametric, model-based approach known as statistical parametric speech synthesis \citep{yoshimura_PhD,Zen_SPSS_SPECOM}.
The concatenative approach builds up the utterance from units of recorded speech, whereas the statistical parametric approach uses a generative model to synthesize the speech.
The statistical parametric approach first extracts a sequence of vocoder parameters \citep{Vocoder} $\bo = \{ \bo_1, \dots, \bo_N \}$ from speech signals $\bx = \{ x_1, \dots, x_T \}$ and linguistic features $\bl$ from the text $W$, where $N$ and $T$ correspond to the numbers of vocoder parameter vectors and speech signals.
Typically a vocoder parameter vector $\bo_n$ is extracted at every 5 milliseconds.
It often includes cepstra \citep{UELS} or line spectral pairs \citep{LSP}, which represent 
vocal tract transfer function, and fundamental
frequency ($F_0$) and aperiodicity \citep{Kawahara_STRAIGHT_Excitation}, which represent characteristics of vocal source excitation signals.
Then a set of generative models, such as hidden Markov models (HMMs) \citep{yoshimura_PhD}, feed-forward neural networks \citep{Zen_DNN_ICASSP}, and recurrent neural networks \citep{Robinson_RNNTTS,Karaani_RTDNNTTS,Fan_BLSTM_Interspeech14}, is trained from the extracted vocoder parameters and linguistic features as
\begin{equation}
  \hat{\Lambda} = \argmax_{\Lambda} p\left( \bo \mid \bl, \Lambda \right),
\end{equation}
where $\Lambda$ denotes the set of parameters of the generative model.
At the synthesis stage, the most probable vocoder parameters are generated given linguistic features extracted from a text to be synthesized as
\begin{equation}
  \hat{\bo} = \argmax_{\bo} p ( \bo \mid \bl, \hat{\Lambda} ).
\end{equation}
Then a speech waveform is reconstructed from $\hat{\bo}$ using a vocoder.
The statistical parametric approach offers various advantages over the
concatenative one such as small footprint
and flexibility to change its voice characteristics.
However, its subjective naturalness is often significantly worse than that of the concatenative approach; synthesized speech often sounds muffled and has artifacts.
\cite{Zen_SPSS_SPECOM} reported three major factors that can degrade the subjective naturalness; quality of vocoders, accuracy of generative models, and effect of oversmoothing.
The first factor causes the artifacts and the second and third factors lead to the muffleness in the synthesized speech.
There have been a number of attempts to address these issues individually, such as developing high-quality vocoders \citep{Kawahara_STRAIGHT,Vocaine,WORLD}, improving the accuracy of generative models \citep{Zen_trjHMM_CSL,Zen_DNN_ICASSP,Fan_BLSTM_Interspeech14,Uria_TrajectoryRNADE_ICASSP2015}, and compensating the oversmoothing effect \citep{Toda_MLGV_IEICE,Takamichi_ModulationSpectrum_TASLP}.
\cite{Zen_LSTMprod_Interspeech} showed that state-of-the-art statistical parametric speech syntheziers matched state-of-the-art concatenative ones in some languages.
However, its vocoded sound quality is still a major issue.

Extracting vocoder parameters can be viewed as estimation of a generative model parameters given speech signals \citep{LPC,UELS}.
For example, linear predictive analysis \citep{LPC}, which has been used in speech coding, assumes that the generative model of speech signals is 
a linear auto-regressive (AR) zero-mean Gaussian process;
\begin{align}
  x_t &= \sum_{p=1}^P a_p x_{t-p} + \epsilon_t \\
  \epsilon_t &\sim \Gauss(0, G^2)
\end{align}
where $a_p$ is a $p$-th order linear predictive coefficient (LPC) and $G^2$ 
is a variance of modeling error.
These parameters are estimated based on the maximum likelihood (ML) criterion.
In this sense, the training part of the statistical parametric approach can be viewed as a two-step optimization and sub-optimal: extract vocoder parameters by fitting a generative model of speech signals then model trajectories of the extracted vocoder parameters by a separate generative model for time series \citep{Tokuda_ASRU2011}.
There have been attempts to integrate these two steps into a single one
\citep{STAVOCO,MGELSD,Maia_WaveformModel_SSW,Kazuhiro_McepHMM_IEICE,Black_AutoEncoder_arXiv,Tokuda_CepLSTM_ICASSP2015,Tokuda_MixCepLSTM_ICASSP2016,Takaki_FFTDNN_ICASSP}.
For example, \cite{Tokuda_MixCepLSTM_ICASSP2016} integrated non-stationary, nonzero-mean Gaussian process generative model of speech signals and LSTM-RNN-based sequence generative model to a single one and jointly optimized them by back-propagation.
Although they showed that this model could approximate natural speech signals, its segmental naturalness was significantly worse than the non-integrated model due to over-generalization and over-estimation of noise components in speech signals.

The conventional generative models of raw audio signals have a number of assumptions which are inspired from the speech production, such as
\begin{itemize}
    \item Use of fixed-length analysis window; They are typically based on a stationary stochastic process \citep{LPC,UELS,Poritz_ARHMM_ICASSP82,Juang_MARHMM_ICASSP85,Kameoka_MultiKernelLPC_ASJ}.  To model time-varying speech signals by a stationary stochastic process, parameters of these generative models are estimated within a fixed-length, overlapping and shifting analysis window (typically its length is 20 to 30 milliseconds, and shift is 5 to 10 milliseconds).
    However, some phones such as stops are time-limited by less than 20 milliseconds \citep{Rabiner_ASR}.
    Therefore, using such fixed-size analysis window has limitations.
    \item Linear filter; These generative models are typically realized as a linear time-invariant filter  \citep{LPC,UELS,Poritz_ARHMM_ICASSP82,Juang_MARHMM_ICASSP85,Kameoka_MultiKernelLPC_ASJ} within a windowed frame.  However, the relationship between successive audio samples can be highly non-linear.
    \item Gaussian process assumption; The conventional generative models are based on Gaussian process \citep{LPC,UELS,Poritz_ARHMM_ICASSP82,Juang_MARHMM_ICASSP85,Kameoka_MultiKernelLPC_ASJ,Tokuda_CepLSTM_ICASSP2015,Tokuda_MixCepLSTM_ICASSP2016}.  From the source-filter model of speech production \citep{Chiba_SourceFilter,Fant_SourceFilter} point of view, this is equivalent to assuming that a vocal source excitation signal is a sample from a Gaussian distribution \citep{LPC,UELS,Poritz_ARHMM_ICASSP82,Juang_MARHMM_ICASSP85,Tokuda_CepLSTM_ICASSP2015,Kameoka_MultiKernelLPC_ASJ,Tokuda_MixCepLSTM_ICASSP2016}. 
    Together with the linear assumption above, it results in assuming that speech signals are normally distributed.
    However, distributions of real speech signals can be significantly different from Gaussian.
\end{itemize}
Although these assumptions are convenient, samples from these generative models tend to be noisy and lose important details to make these audio signals sounding natural.

WaveNet, which was described in Section~\ref{sec:wavenet}, has none of the above-mentioned assumptions.
It incorporates almost no prior knowledge about audio signals, except the choice of the receptive field and $\mu$-law encoding of the signal.
It can also be viewed as a non-linear causal filter for quantized signals.
Although such non-linear filter can represent complicated signals while preserving the details, designing such filters is usually difficult \citep{NonlinearFilterDesign}.
WaveNets give a way to train them from data.

\section{Details of TTS Experiment}
\label{appendix:tts_experiment}

The HMM-driven unit selection and WaveNet TTS systems were built from speech at 16 kHz sampling.
Although LSTM-RNNs were trained from speech at 22.05 kHz sampling, 
speech at 16 kHz sampling was synthesized at runtime using a resampling functionality in the Vocaine vocoder \citep{Vocaine}.
Both the LSTM-RNN-based statistical parametric and HMM-driven unit selection speech synthesizers were built from the speech datasets in the 16-bit linear PCM, whereas the WaveNet-based ones were trained from the same speech datasets in the 8-bit $\mu$-law encoding.

The linguistic features include phone, syllable, word, phrase, and utterance-level features \citep{HTS_label} (\eg phone identities, syllable stress, the number of syllables in a word, and position of the current syllable
in a phrase) with additional frame position and phone duration features \citep{Zen_DNN_ICASSP}. These features were derived and associated with speech every 5 milliseconds by phone-level forced alignment at the training stage.
We used LSTM-RNN-based phone duration and autoregressive CNN-based $\log F_0$ prediction models.
They were trained so as to minimize the mean squared errors (MSE).
It is important to note that no post-processing was applied to the audio signals generated from the WaveNets.

The subjective listening tests were blind  and crowdsourced. 
100 sentences not included in the training data were used for evaluation.
Each subject could evaluate up to 8 and 63 stimuli for North American English and Mandarin Chinese, respectively.
Test stimuli were randomly chosen and presented for each subject.
In the paired comparison test, each pair of speech samples was the same text synthesized by the different models.
In the MOS test, each stimulus was presented to subjects in isolation.
Each pair was evaluated by eight subjects in the paired comparison test, and each stimulus was evaluated by eight subjects in the MOS test.
The subjects were paid and native speakers performing the task.
Those ratings (about 40\%) where headphones were not used were excluded when computing the preference and mean opinion scores.
\tblref{tab:sxs} shows the full details of the paired comparison test shown in \figref{fig:sxs2}.

\begin{table}[htbp]
    \centering
    \begin{tabularx}{0.9\textwidth}{L|RRRRR|R}
    \toprule
                       & \multicolumn{5}{|c|}{\textbf{Subjective preference (\%) in naturalness}} &  \\ \cmidrule{2-6}
                       & & & \textbf{WaveNet} & \textbf{WaveNet} & No & \\
     \textbf{Language} & \textbf{LSTM} & \textbf{Concat} & (L) & (L+F) & preference & $p$ value \\ \midrule\midrule
     North    & 23.3 & \textbf{63.6} & & & 13.1 & $\ll 10^{-9}$ \\
     American & 18.7 & & \textbf{69.3} & & 12.0 & $\ll 10^{-9}$ \\
     English  & 7.6 & & & \textbf{82.0} & 10.4 & $\ll 10^{-9}$ \\
              & & 32.4 & \textbf{41.2} & & 26.4 & $0.003$ \\
              & & 20.1 & & \textbf{49.3} & 30.6 & $\ll 10^{-9}$ \\
              & & & 17.8 & \textbf{37.9} & 44.3 & $\ll 10^{-9}$ \\
     \midrule
     Mandarin & \textbf{50.6} & 15.6 & & & 33.8 & $\ll 10^{-9}$ \\
     Chinese  & 25.0 & & 23.3 & & 51.8 & 0.476 \\
              & 12.5 & & & \textbf{29.3} & 58.2 & $\ll 10^{-9}$ \\
              & & 17.6 & \textbf{43.1} & & 39.3 & $\ll 10^{-9}$ \\
              & & 7.6 & & \textbf{55.9} & 36.5 &  $\ll 10^{-9}$ \\
              & & & 10.0 & \textbf{25.5} & 64.5 &  $\ll 10^{-9}$ \\
     \bottomrule
   \end{tabularx}%
   \caption{Subjective preference scores of speech samples between LSTM-RNN-based statistical parametric (\textbf{LSTM}), HMM-driven unit selection concatenative (\textbf{Concat}), and proposed WaveNet-based speech synthesizers.
   Each row of the table denotes scores of a paired comparison test between two synthesizers.
   Scores of the synthesizers which were significantly better than their competing ones at $p < 0.01$ level were shown in the bold type. 
   Note that \textbf{WaveNet} (L) and \textbf{WaveNet} (L+F) correspond to WaveNet conditioned on linguistic features only and that conditioned on both linguistic features and $F_0$ values. }
   \label{tab:sxs}
\end{table}%

\end{document}